\documentclass[%
 aip,
 rsi,%
 amsmath,amssymb,
 reprint,%
 article-title,
]{revtex4-1}

\usepackage{graphicx}
\usepackage{dcolumn}
\usepackage{bm}
\usepackage[breaklinks=true,colorlinks=true,linkcolor=blue,urlcolor=blue,citecolor=blue]{hyperref}
\usepackage[]{units}	
\usepackage{textcomp}
\usepackage[all]{hypcap}
\usepackage{xcolor}

\begin{document}

\preprint{AIP/123-QED}

\title[]{Fast, precise, and widely tunable frequency control of an optical parametric oscillator referenced to a frequency comb}

\author{Alexander Prehn}
\author{Rosa Gl\"ockner}
\author{Gerhard Rempe}
\author{Martin Zeppenfeld}
 \email{Martin.Zeppenfeld@mpq.mpg.de.}
\affiliation{ 
Max-Planck-Institut f\"ur Quantenoptik, Hans-Kopfermann-Str. 1, 85748 Garching, Germany
}


\begin{abstract}
Optical frequency combs (OFC) provide a convenient reference for the frequency stabilization of continuous-wave lasers. We demonstrate a frequency control method relying on tracking over a wide range and stabilizing the beat note between the laser and the OFC. The approach combines fast frequency ramps on a millisecond timescale in the entire mode-hop free tuning range of the laser and precise stabilization to single frequencies. We apply it to a commercially available optical parametric oscillator (OPO) and demonstrate tuning over more than 60\,GHz with a ramping speed up to 3\,GHz/ms. Frequency ramps spanning 15\,GHz are performed in less than 10\,ms, with the OPO instantly relocked to the OFC after the ramp at any desired frequency. The developed control hardware and software is able to stabilize the OPO to sub-MHz precision and to perform sequences of fast frequency ramps automatically.
\end{abstract}

\maketitle

\section{\label{sec:intro}Introduction}

Modern experiments in atomic and molecular physics often use widely tunable lasers and require both large tunability and precise frequency stabilization. In many cases, spectroscopy has to be performed with stable lasers in a large frequency range to find and characterize initially unknown atomic or molecular states, for example for the production of molecules from ultracold atoms\cite{Ni2008,Park2015} or magneto-optical trapping of molecules\cite{Norrgard2016,Devlin2015}. Furthermore, it can be beneficial to address different known transitions spaced several GHz or even tens of GHz sequentially with a single laser. For example, we use a mid-infrared optical parametric oscillator (OPO) to excite a number of rovibrational transitions spaced more than 10\,GHz in cold molecules held in an electrostatic trap\cite{Englert2011}. Frequency switching on a millisecond timescale allows us to perform motional cooling, rotational-state preparation, and state detection with a single laser in the same run of an experiment\cite{Glockner2015a,Glockner2015}. 

Self-referenced optical frequency combs (OFC)\cite{Diddams2000} which are now available commercially have become a common tool for stabilizing lasers in a wide bandwidth by stabilizing the radiofrequency beat note between the laser and a mode of the OFC. A number of methods have been developed to achieve both absolute frequency stability and large tunability over many comb modes of a continuous-wave (cw) laser or OPO referenced to an OFC. For tuning, the comb\cite{Park2006,Peltola2014}, the cw laser\cite{Jost2002,Schibli2005,Benkler2013a,Fordell2014,Gunton2015}, or both\cite{Ahtee2009} can be scanned. Tuning over 10\,GHz without changing the frequency lock has been demonstrated by adding an external electro-optic modulator\cite{Inaba2006}. Further, phase-stable tuning over almost 30\,GHz in about a second has been achieved by shifting the carrier-envelope offset frequency of the OFC between subsequent pulses of the OFC\cite{Benkler2013,Rohde2014}.

The objective of this work was to build a device tuning our frequency-comb-referenced OPO over tens of GHz in a couple of milliseconds, yet stabilizing the OPO to sub-MHz precision on single frequencies. These requirements were set because we wanted to drive several rovibrational transitions in polyatomic molecules quasi-simultaneously for optical pumping, demanding frequency switching in a time shorter than the typical decay time of vibrational excitations. 
Additionally, the OFC's spectrum should remain unchanged during frequency tuning of the OPO such that several independent lasers can be stabilized to the same OFC.

In this paper, we demonstrate fast, precise and widely tunable frequency control of the idler wave of a singly-resonant OPO by controlling the frequencies of pump and signal which are referenced to the OFC. We show tuning over more than \unit[60]{GHz}, with \unit[15]{GHz} ramps performed in less than \unit[10]{ms}. Although not actively stabilized during a ramp, the OPO frequencies are instantly relocked to the OFC at the end of each ramp. Sequences of fast ramps to any idler frequency within the mode-hop free tuning range can be performed hands-free and reliably.

\section{\label{sec:basics}Basic principles}

\begin{figure}[tb]
\centering
\includegraphics{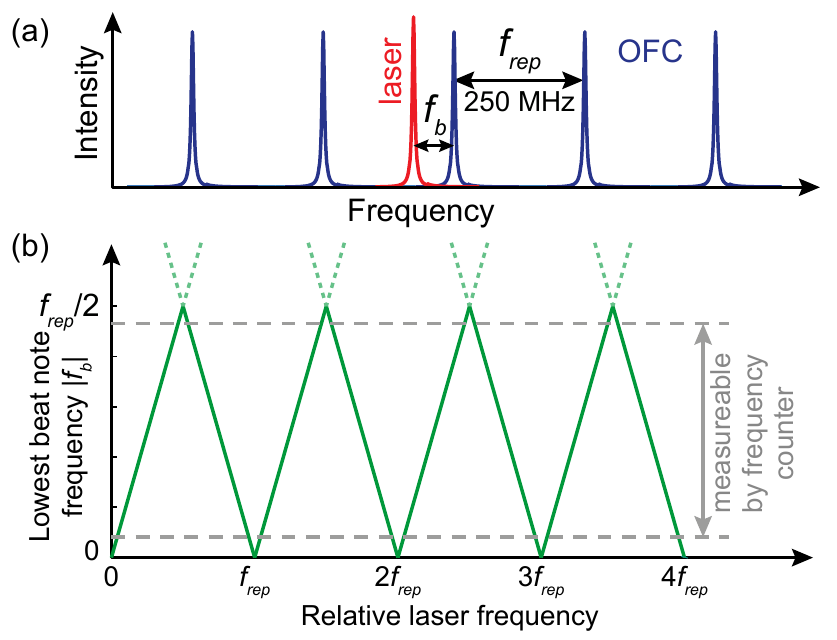}
\caption{
Tracking the laser frequency with a stationary optical frequency comb (OFC). (a)~Spectrum of frequencies of a laser beam and a few comb modes. (b)~Beat note frequency versus laser frequency during tuning. The lowest beat note frequency changes in a regular tooth structure defined by $f_{rep}$ when the laser frequency is ramped relative to the frequency of a certain comb mode. If the laser frequency lies about midway between two comb modes, the lowest and next to lowest beat frequencies are close to each other which is indicated by the dashed extensions of the tooth structure.
}
\label{fig:beat}
\end{figure}

The frequency control mechanism is based on beating the laser with an OFC and using the lowest beat note for frequency tracking during tuning and for precise stabilization. The principle is illustrated in Fig.~\ref{fig:beat}. The lowest ``signed'' beat frequency $f_b$ determines the laser frequency with respect to the nearest comb mode. ``Signed'' refers to $f_b$ being positive (negative) if the closest comb mode has a smaller (larger) optical frequency. $f_b$ is confined by the repetition rate $f_{rep}$ of the mode-locked laser generating the OFC: $-f_{rep}/2\leq f_b\leq f_{rep}/2=125\,\mathrm{MHz}$. Consequently, tuning the frequency of the laser results in a regular tooth structure of $|f_b|$ as shown. We start with the laser being locked to the OFC at a known absolute optical frequency. To perform a frequency ramp, the lock is switched off and the frequency is swept while continuously measuring $f_b$ and counting the comb modes passed. Once the target frequency is reached, the ramp is stopped and the laser is instantly relocked to the OFC by stabilizing $f_b$ at a value of choice. 

\begin{figure}[tb]
\centering
\includegraphics{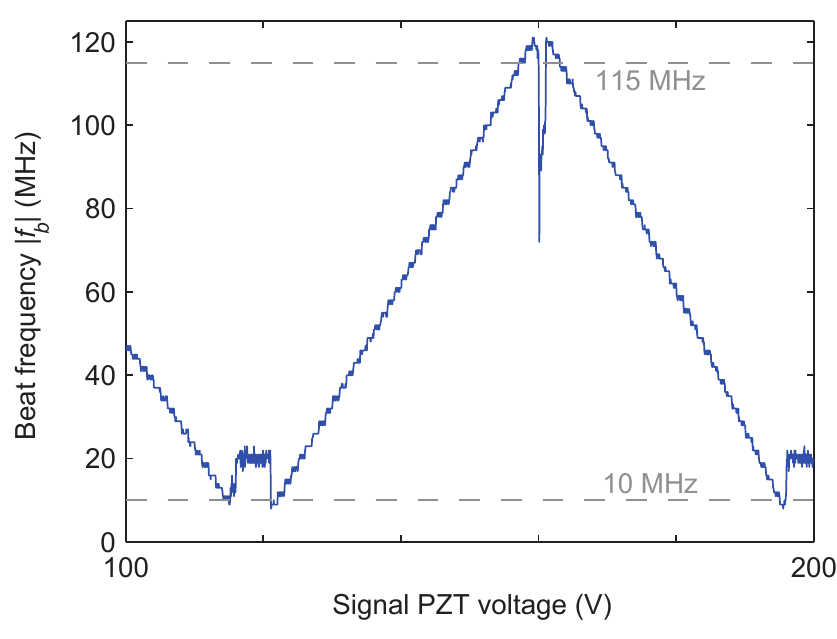}
\caption{
Counted beat frequency during a slow linear ramp of the free-running signal beam. The optical frequency was changed by applying a voltage ramp (steps of \unit[1]{V}) to the piezo-electric transducer (PZT) of the OPO cavity (see Sec.~\ref{sec:setup}). The dashed lines mark the \unit[-3]{dB} points of the radio-frequency filters. Outside the pass band the frequency counter measures noise.
}
\label{fig:signalramp}
\end{figure}

The main difficulty of controlling the beat frequency during tuning over many comb lines lies in correct handling of the frequency ranges in which the laser frequency coincides with a comb mode or lies midway between two modes. In the former case $f_b$ is zero, in the latter the lowest and next higher beat frequencies are degenerate and can be mixed up (see Fig.~\ref{fig:beat}b). Demonstrated solutions to this problem include sudden jumps to a beat note of opposite sign via a sudden step of a control voltage\cite{Fordell2014} or the use of an acousto-optic modulator\cite{Schibli2005,Gunton2015} to avoid such regions of the spectrum. In both cases, a lock to the OFC was maintained during frequency tuning~\cite{Fordell2014,Schibli2005,Gunton2015} possibly limiting the speed. Our frequency tracking approach allows for fast ramping to a target frequency independent of the locking electronics. We can simply ignore frequency measurements in the delicate parts of the spectrum and interpolate the frequency there without losing track of it. To ensure sufficient suppression of beat frequencies of higher order near $|f_b|=f_{rep}/2$ and block low-frequency noise we chose to filter the beat note with a pass band of 10 to 115 MHz (\unit[-3]{dB} points of filters). This permits clean measurements of $f_b$ inside the pass band. An example of a measurement for the signal wave of the OPO during a linear ramp is shown in Fig.~\ref{fig:signalramp}. The data was directly obtained with the control electronics (see Sec.~\ref{sec:electronics}) and its quality already suggests that tracking of the optical frequency will be possible.

Due to the unavoidable filtering, the parts of the beat spectrum outside the pass band of the filters are not directly accessible to frequency measurement and locking. If one wanted to control the frequency of a single laser beam and demanded locking of the laser to an arbitrary optical frequency inside the tuning range, an additional frequency shifter, e.g., an electro- or acousto-optic modulator, would have to be introduced\cite{Schibli2005,Ricciardi2012,Gunton2015}. It would shift the frequency by, e.g., 20\,MHz if the desired lock frequency lies in the clipped areas of the beat spectrum. We note that this shift would have to be applied once per ramp, independent of the number of comb modes crossed for that ramp.

We apply the frequency control scheme to the idler wave of a singly-resonant OPO (see Sec.~\ref{sec:setup} and Fig.~\ref{fig:opo}). In our case the aforementioned filtering does not pose any restrictions. The idler frequency $\nu_i = \nu_p - \nu_s$ can be changed by independently tuning the pump laser ($\nu_p$) or the signal frequency $\nu_s$ which is resonant to the OPO cavity and we reference the latter two to the OFC. The optical frequency of each comb mode $N$ of a self-referenced OFC is $\nu_N = N f_{rep} + f_{ceo}$ with the carrier-envelope offset frequency $f_{ceo}$. Then, we find $\nu_{p/s}=Nf_{rep}+f_{ceo}+f_{b,p/s}$, where $f_{b,p}$ and $f_{b,s}$ are the signed beat frequencies of pump and signal with the comb. Thus, the idler frequency can be expressed as
\[
\nu_i = N'f_{rep} + f_{b,p} - f_{b,s}.
\]
Note that $f_{ceo}$ cancels here. For reasons which will be explained later, the final lock point of the pump beam $f_{b,p}$ is always set to be well separated from the filter cut-off, $50\,\mathrm{MHz}\leq|f_{b,p}|\leq90\,\mathrm{MHz}$. The idler frequency can then be ramped to arbitrary values without employing additional frequency shifters by choosing $f_{b,s}$ appropriately inside the pass band of the filters.

\section{\label{sec:setup}Experimental setup}

\begin{figure}[tb]
\centering
\includegraphics{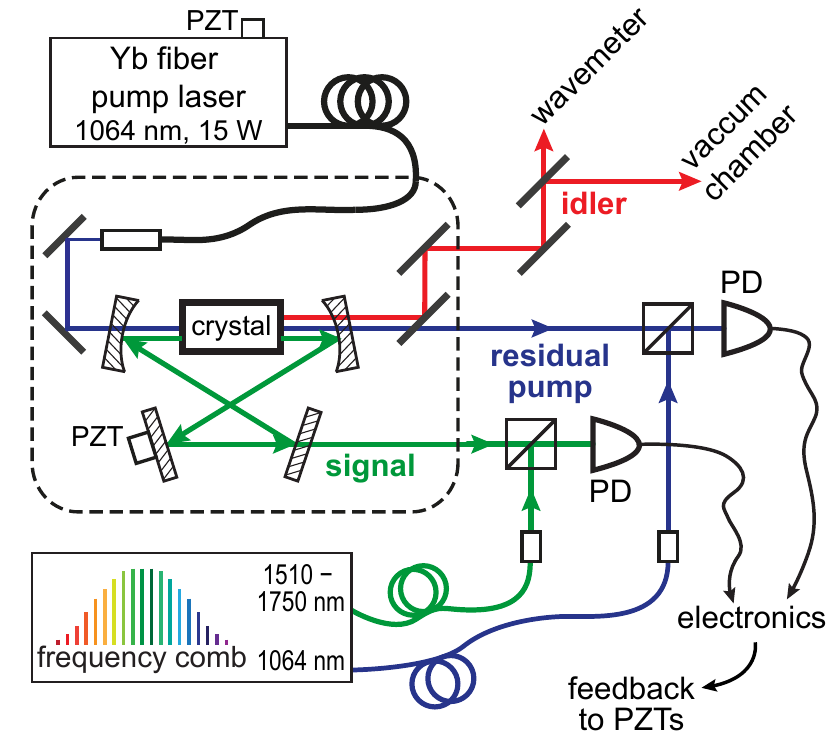}
\caption{
Schematic of the OPO system and the frequency detection setup. Signal and idler beams are generated in a singly-resonant OPO pumped by a fiber laser. The absolute frequency of the idler beam is determined with a wavemeter. Pump and signal beams are mixed with light from an optical frequency comb to allow precise and fast relative measurements of their frequencies via detection of the beat note with fast photodiodes (PD). The beat note frequencies are processed by the control electronics which provides feedback to the PZT of the pump laser and the OPO cavity.
}
\label{fig:opo}
\end{figure}

A simple sketch of the optical part of the experimental apparatus is shown in Fig.~\ref{fig:opo}. The commercially available cw OPO (Lockheed Martin Aculight, Argos 2400-SF-15, module C), which has been described in detail elsewhere\cite{Henderson2006,Henderson2006a,Morrison2013}, is pumped with up to \unit[15]{W} at \unit[1064]{nm} and generates an idler wave in the range of \unit[3.2]{} to \unit[3.9]{\textmu m}. Coarse wavelength tuning of idler and signal is realized by varying the position of the periodically poled nonlinear crystal and the tilt angle of an intracavity etalon with a free spectral range of \unit[400]{GHz}. Although it has been shown before that both elements can be tuned with a computer-controlled algorithm\cite{Morrison2013}, for us a manual adjustment of those two components suffices. A piezo-electric transducer (PZT) varying the cavity length allows mode-hop-free tuning of the signal frequency over more than one free spectral range of the cavity which is about \unit[593]{MHz}. The pump frequency can be tuned continuously over \unit[100]{GHz} by strain variation of the seed laser fiber length via another PZT element, which is the main tuning mechanism used to adjust the idler frequency during experiments.

The setup provides two complementary modes of frequency measurement. First, fast and precise frequency measurements relative to a known initial frequency are obtained with the OFC. Therefore, we beat a few mW of pump and signal light with radiation from a self-referenced OFC synthesizer (Menlo Systems, FC1500), which provides a frequency comb with $f_{rep}=250\,\mathrm{MHz}$ mode spacing at \unit[1064]{nm} and about 1510 to \unit[1750]{nm}. Repetition rate and offset frequency are stabilized to a stable \unit[10]{MHz} reference obtained from an H$_2$-Maser. The resulting beat notes of pump and signal beams are measured with high speed InGaAs photodiodes (Thorlabs, DET01CFC) and processed further by the electronics. Second, the idler frequency can be determined with an absolute accuracy of $\pm 20\,\mathrm{MHz}$ with a calibrated wavemeter (Bristol Instruments, 621A-IR). However, wavemeter measurements can be performed with a maximum rate of \unit[2.5]{Hz} and after large frequency ramps the device needs up to about \unit[2]{s} to adjust to the new frequency and display correct results. The wavemeter is used to fix the absolute optical frequency once before fast ramps can be performed.

\section{\label{sec:electronics}Control electronics}

\begin{figure*}[tbp]
\centering
\includegraphics{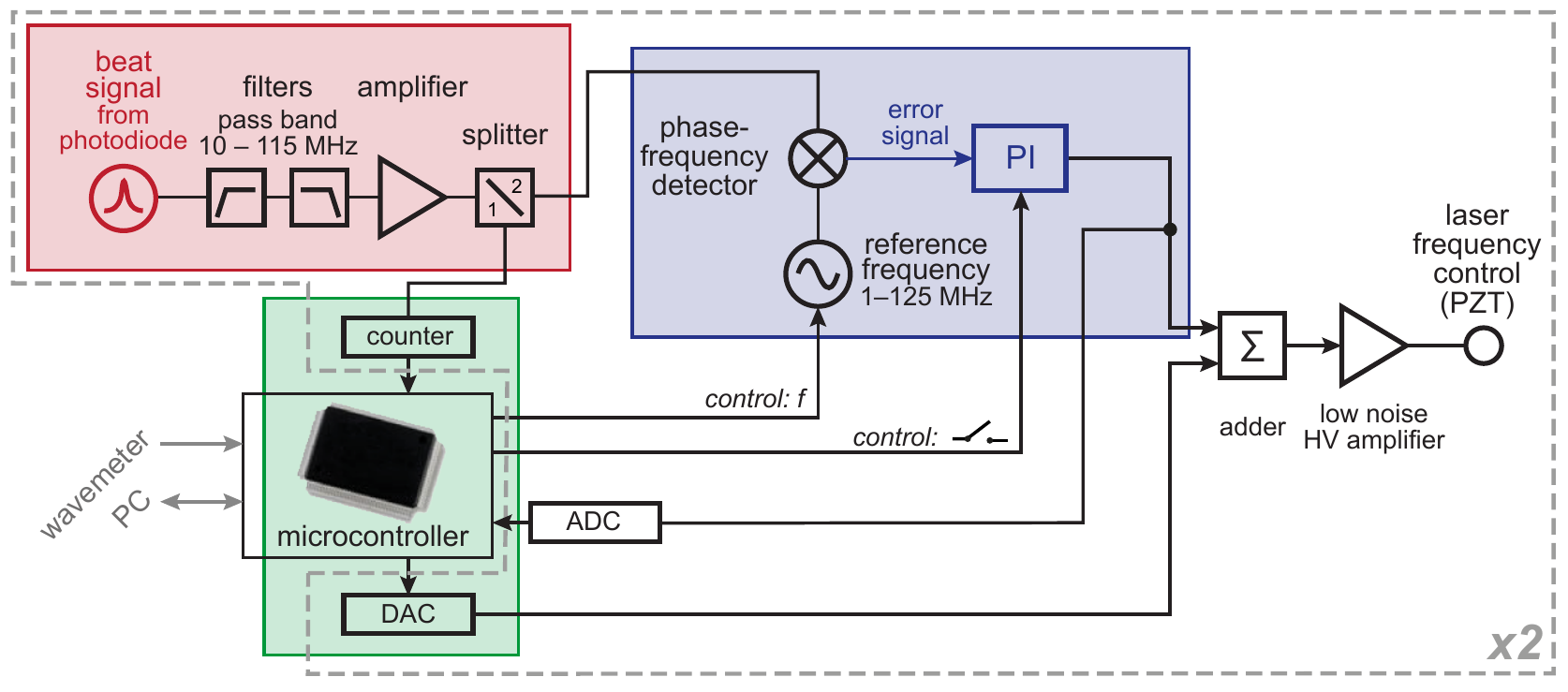}
\caption{
Schematic of the electronics consisting of three main parts: Beat signal processing (red), digitally controlled fast frequency ramping (green), and an analog beat lock circuit (blue). The microcontroller measures and controls all relevant parameters. All parts enclosed by the dashed line exist twice as we control two optical frequencies. The operation of the electronics and the single components are described in the main text.
}
\label{fig:electronics}
\end{figure*}

The structure of the custom-built electronics controlling the two beat frequencies and hence the optical frequencies of the OPO reflects the fact that the OPO system can basically be in two states: performing a frequency ramp or being locked to frequencies of choice. Ramps are controlled by a microcontroller that measures and adjusts all relevant parameters. The microcontroller is an Atmel AT91SAM7XC256 with 32\,bit ARM-based architecture which runs custom software programmed in C. The software is specialized to our OPO and our experimental sequences but it is available from the authors upon request. For precise stabilization to a single frequency between two ramps we use an analog proportional-integral (PI) regulation circuit\cite{Bechhoefer2005}. Except for the microcontroller all components exist twice as we have to control the pump and the signal beam of the OPO.

A schematic of the electronics is displayed in Fig.~\ref{fig:electronics}. The beat signal recorded by the photo diode is filtered (pass band \unit[10--115]{MHz}) and amplified (red box in the figure). For frequency ramping and tracking (green box) the beat signal is digitized by a frequency counter with a high sampling rate of \unit[62.5]{kHz}. Based on the frequency measurement the microcontroller applies voltage ramps to the PZT element of the laser (via a high precision, low-noise DAC\footnote{Analog Devices AD5791} and a fast, low-noise high voltage amplifier\footnote{Falco Systems WMA-280, \unit[0--150]{V}, for the pump laser, Trek Model 2210, \unit[0--1000]{V}, for the signal cavity}) and thus ramps the laser frequency. 

Once the laser is at the desired frequency, the PI regulation circuit stabilizes the beat frequency (blue box in schematic). A reference frequency in the range of \unit[1--125]{MHz} is generated via Direct Digital Synthesis\footnote{Analog Devices AD9959 evaluation board}. Comparing the recorded beat signal and the reference, a custom-built phase-frequency detector\footnote{Built around the chip MCH12140} produces an error signal that is fed to the PI controller. The output voltage of the PI controller is scaled such that the frequency range covered by the PI regulation loop is rather small, on the order of \unit[10]{MHz}. 

The total voltage applied to the PZTs via the high voltage amplifiers is a sum of two contributions, the output of the DAC set by the microcontroller and the output of the PI controller. Consequently, for both pump and signal the microcontroller sets the optical frequency approximately by adjusting the DAC voltage during a ramp, which can span many GHz. On top of that, precise frequency stabilization is realized with the analog control loop. When the laser is locked to a fixed frequency, the microcontroller measures the output voltage of the PI controller with a built-in ADC and adjusts the DAC voltage if the PI voltage approaches a border of the regulation range due to long-term drifts. Note that the microcontroller also controls the reference frequency for the beat lock, can switch the PI controller on and off and receives measurements from the wavemeter. Thus, it is capable of controlling the frequencies of the OPO fully automatized---during ramps and while being locked to the OFC.

\section{\label{sec:freqramps}Frequency ramps}

In this section, we explain in detail the implementation of fast frequency ramps of the idler starting with the OPO being locked to the OFC at a known absolute frequency $\nu_i = Nf_{rep} + f_{b,p} - f_{b,s}$ (cf. Sec.~\ref{sec:basics}). As mentioned before, the only optical frequency of interest for us is the idler frequency. Furthermore, idler tuning is mainly accomplished by tuning the pump laser, whereas the signal frequency is ramped only for fine tuning. For internal bookkeeping, we therefore process the frequencies in a simplified manner as $\nu_i = f_p - f_s$, with the contributions from the pump $f_p = N'f_{rep} + f_{b,p}$ and from the signal beam $f_s = f_{b,s}$. 

Our basic protocol for a frequency ramp is the following. First, the target frequencies (lock frequencies) $f'_p$ and $f'_s$ are calculated with the boundary conditions $50\,\mathrm{MHz}\leq|f'_{b,p}|\leq90\,\mathrm{MHz}$ and $10\,\mathrm{MHz}\leq|f'_{b,s}|\leq115\,\mathrm{MHz}$. The latter condition is just given by the pass band of the radio-frequency filters. Note again that these conditions do not restrict the idler frequency to specific values. Second, the PI controllers are switched off, i.e., the lock to the OFC is released. In fact, we only switch off the integral parts for technical reasons. This ensures that the controllers are in well-defined states at the time of relock. Third, the reference frequencies for the final beat lock are set. Fourth, the optical frequencies of pump and signal are ramped towards the target values, and $f_p$ and $f_s$ are tracked by repeatedly measuring the beat frequencies. The voltage applied to the PZTs and the frequency counter measurements are updated every \unit[16]{\textmu s}. Finally, the OPO is relocked to the OFC as soon as the target frequencies are reached by reactivating the PI controllers.

Special attention has to be paid to the frequency ramping and tracking part for two reasons. First, the tracking algorithm has to discard beat frequency measurements in the clipped areas of the beat spectrum (see Fig.~\ref{fig:signalramp}) and interpolate the frequency there. Second, many piezo-electric crystals show a nonlinear response to an applied voltage ramp. In particular, for the PZT in our pump laser we observe a delayed response of the piezo to an applied voltage and drifts of the laser frequency after the end of a fast voltage ramp. Even for a fixed step size, the delay and further drift depend on the direction of the voltage ramp, the amplitude, the slew rate, and the starting voltage. This precludes the use of any kind of look-up table relating an applied voltage to a particular frequency and is one of the main reasons for implementing the frequency tracking with ``live'' feedback. 

Frequency tracking is identical for both pump and signal. In every step of the ramp (every \unit[16]{\textmu s}) a predicted laser frequency $f_{pred,i}$ and a predicted frequency change for the next step $\Delta f_{pred,i+1}$ are calculated as follows. A preliminary value $\tilde{f}_{pred,i}$ is obtained from $\tilde{f}_{pred,i} = f_{pred,i-1} + \Delta f_{pred,i}$. To compare the predicted value with measurement, the signed beat frequency $f_{b,pred}$ is computed from $\tilde{f}_{pred,i}$\footnote{In particular, $f_{mod} = \tilde{f}_{pred,i}$ modulo $f_{rep}$. $f_{b,pred} = f_{mod}$, if $f_{mod}\leq f_{rep}/2$; $f_{b,pred} = f_{mod}-f_{rep}$, if $f_{mod}> f_{rep}/2$.}.
Only if $f_{b,pred}$ lies in a frequency range where good counter measurements are expected, i.e., in the pass band of the filters, we calculate the deviation of the measured value $\left| f_b \right|$ from the predicted frequency as
\begin{align*}
f_{corr} &= \left| f_b \right| - f_{b,pred}, &\mathrm{if}\ f_{b,pred}>0, \\
f_{corr} &= - \left| f_b \right| - f_{b,pred}, &\mathrm{if}\ f_{b,pred}<0.
\end{align*}
Finally, the two parameters of interest are corrected as
\begin{align*}
f_{pred,i} &= \tilde{f}_{pred,i} + a \cdot f_{corr} \\
\Delta f_{pred,i+1} &= \Delta f_{pred,i} + b \cdot f_{corr},
\end{align*}
with numerical factors $a$ and $b$. To be less sensitive to single measurement errors we continuously average over a few consecutive steps of the calculation by choosing these factors smaller than unity. As a result, this frequency tracking approach is quite insensitive to single corrupted measurements. Moreover, it requires only a few data points per slope of the beat frequency tooth structure to work properly.

We perform all ramps starting with a fixed voltage change per step. Consequently, the initial values for the expected frequency change $\Delta f_{pred,0}$ are not changing and can be predetermined experimentally for each of the two beams. 

The ramp speed and the frequency span, however, differ quite significantly for pump and signal. Ramps of the signal frequency are always much shorter than \unit[500]{MHz}, as they are used for fine tuning only. Consequently, the duration of a signal ramp does not limit the overall performance of the system and we can ramp with relatively low speed. We chose a constant step size of 1\,V/16\,\textmu s which translates to a ramp speed of about 188\,MHz/ms. For these settings we do not observe any significant delays in response or further drifts of the frequency after the end of the ramp. The signal can be directly locked to the OFC by switching on the PI controller, if the measured beat frequency is close to the target frequency. 
In contrast, the ramp has to be optimized more carefully for the pump beam, because ramps span many GHz, the response of the PZT has to be taken into account, and speed matters in this case as it sets the timescale of the entire frequency ramp. We start the ramp with a step size of 0.1\,V/16\textmu s which corresponds to about 3\,GHz/ms in the middle of a frequency ramp. To prevent an overshoot, we significantly slow down the ramp at its end and regulate the applied voltage.

\begin{figure*}[tbp]
\centering
\includegraphics{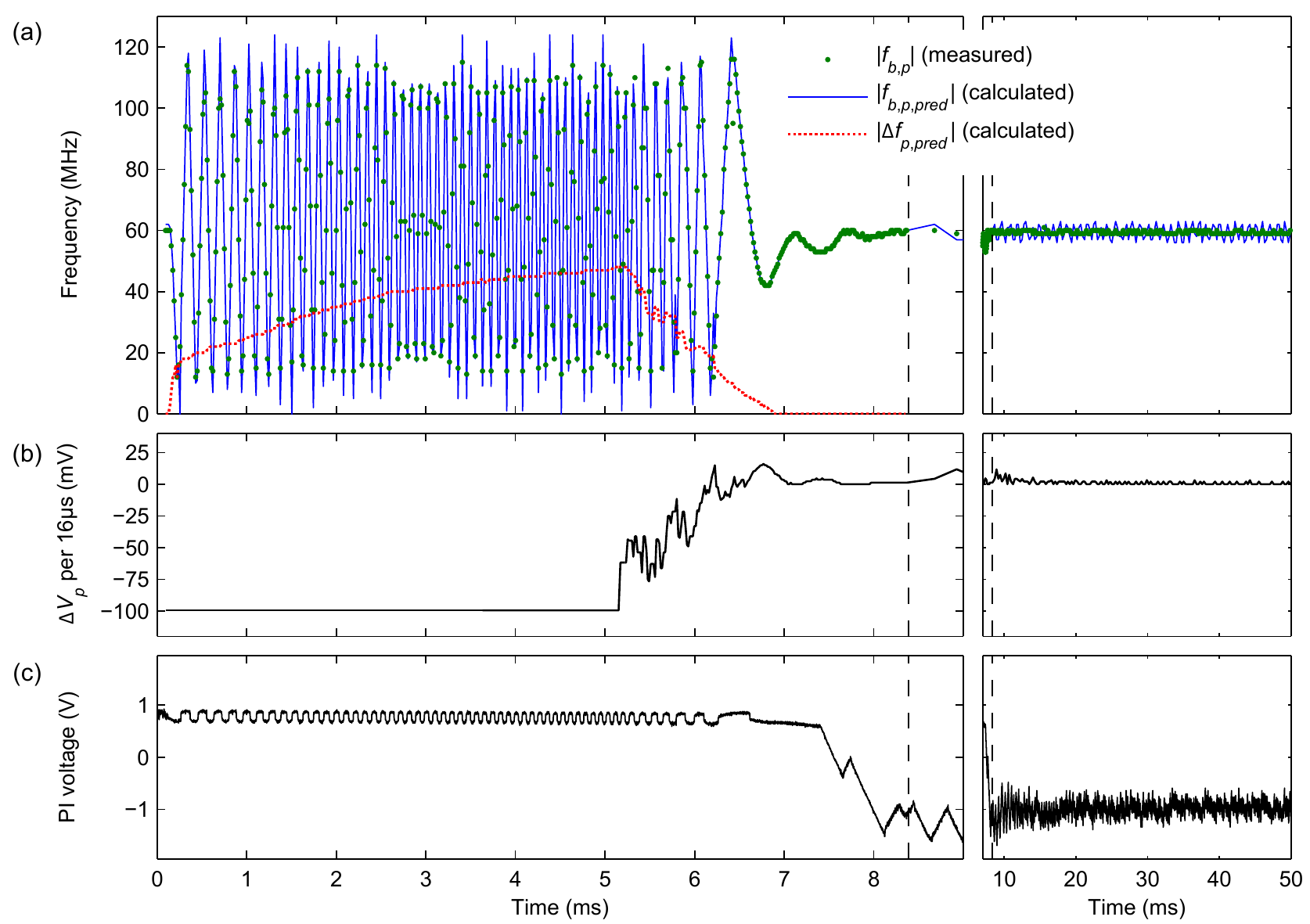}
\caption{
Detailed analysis of a frequency ramp of the pump laser. The idler frequency was changed from \unit[88935777]{MHz} to \unit[88922117]{MHz} in \unit[8.4]{ms} by tuning the frequencies of pump and signal. The left panels show data recorded during the ramp, the panels on the right show data obtained after the completion of the frequency ramp. The vertical dashed line marks the time of reactivation of the PI controller. (a)~Evolution of the laser frequency. The points show the measured beat frequency $\left| f_{b,p} \right|$ which cycles 110 times between 0 and 125\,MHz corresponding to a frequency ramp spanning almost 14\,GHz. The blue solid line represents the absolute laser frequency calculated by the microcontroller and converted to units of the beat frequency. The red dotted line shows the predicted frequency change per step $\left|\Delta f_{p,pred}\right|$. (b)~Actual voltage change per step applied to the PZT. (c)~Output voltage of PI controller. Note that the voltage range available for regulation is $[-15,15]$\,V.
}
\label{fig:rampanalysis}
\end{figure*}

The slowdown of the pump ramp and the regulation at its end have two contributions, ensuring effective compensation of the delayed response and further drifts of the PZT and hence the optical frequency. Both are applied once the target frequency is closer than 2\,GHz. First, the voltage change $\Delta V$ applied to the PZT per step is decreased from initially 0.1\,V to zero approximately proportional to the square root of the frequency difference to the target. This would ideally lead to a linear decrease of ramp speed in time, if the PZT would not show hysteresis. To compensate the overshoot of the PZT we additionally apply a correction to the voltage change $\Delta V$ which is proportional to the difference between the expected linear decrease and the actual frequency change $\Delta f_{pred,i}$.

During the very last part of the ramp, when the pump frequency is less than 20\,MHz away from the target, an effective proportional-differential regulation is implemented by adjusting $\Delta V$ as $\Delta V=c\cdot(f'-f)-d\cdot\Delta f_{pred,i}$ with numerical factors $c,d$. As a result, the actual laser frequency $f$ quickly settles to its target value $f'$. This final regulation in the frequency interval of $\pm20$\,MHz around the target frequency is the reason for restricting the target beat frequency of the pump laser to a narrower band than the pass band of the radio-frequency filters (see above). For the pump beam we simply need some ``frequency space'' to slow the ramp down.

The aforementioned details of the frequency ramping procedure are also visible in data recorded with the control electronics. Figure~\ref{fig:rampanalysis} shows a ramp of the pump laser spanning almost 14\,GHz. In part (a) we plot frequency vs. time, in particular the measured beat frequency $\left| f_{b,p} \right|$ and the predicted value $\left|f_{b,p,pred}\right|$ calculated by the tracking algorithm. The agreement is excellent. Additionally, the predicted frequency change per step $\left|\Delta f_{p,pred}\right|$ is shown. From the curves it is evident that the ramp starts slowly due to the delayed response of the PZT, but accelerates quite quickly as a constant $\Delta V$ per step is applied (Fig.~\ref{fig:rampanalysis}(b)). The slowdown of the ramp is also apparent in both the beat frequency tooth structure and $\left|\Delta f_{p,pred}\right|$. During the final regulation towards the target beat frequency of 60\,MHz there are some oscillations visible. We believe that further optimization of the ramp parameters can eliminate the oscillations, but did not find that the effort is necessary at the moment. In the figure, the vertical dashed lines mark the switch-on of the PI controller after 8.4\,ms of ramp time. As expected the beat frequency is stable from that point on. We plot the output voltage of the PI controller during the whole frequency ramp (see Fig.~\ref{fig:rampanalysis}(c)). The fact that the voltage stays close to zero, i.e. in the center of the regulation range of $[-15,15]\,\mathrm{V}$, all the time demonstrates the instant relock of the OPO to the OFC after the ramp. 

\section{\label{sec:absolutefreq}Determination and verification of the absolute laser frequency}

To this point, we only described the control of the idler frequency via fast ramps starting from a known value. In this section, we outline how we fix and determine the OPO frequency initially and verify it during experiments.

For the precise determination of the absolute idler frequency we use the wavemeter in conjunction with the beat frequencies of pump and signal. For the determination of the initial idler frequency $\nu_i = Nf_{rep} + f_{b,p} - f_{b,s}$ the following protocol is employed. Starting from a random point, we slowly ramp the frequencies of pump and signal until the corresponding beat frequencies are equal to $f_{b,p}=+50\,\mathrm{MHz}$ and $f_{b,s}=+30\,\mathrm{MHz}$ and beat-lock the OPO to the OFC. Then, the wavemeter is read out and $Nf_{rep}$ in the above equation is determined from the reading. As $Nf_{rep}$ is a multiple of 250\,MHz in our setup, a wavemeter with an accuracy on the order of 100\,MHz would suffice to determine the idler frequency to sub-MHz precision. From there, fast frequency ramps to any frequency within the tuning range can be performed. This initial frequency determination which is fully automated takes a couple of seconds.

During our experiments with cold molecules we verify the idler frequency at least once per experimental sequence. We therefore reserve a couple of seconds in which no frequency ramps are performed and the OPO is locked to a fixed frequency for the wavemeter readout. In single experimental sequences we often perform a couple of thousand frequency ramps. Checking the absolute frequency only once assumes that the system does not perform multiple ramps with errors of $n\cdot$250\,MHz which exactly cancel each other. However, after months of experiments with more than $10^6$ ramps performed on some days we have no indication that such errors occur in practice. All other errors, e.g., a lock to a beat frequency of wrong sign or a PI controller running at a limit of the regulation range, are detected by the control software or during verification with the wavemeter. In the case of an error, the data obtained in this sequence is discarded, the OPO is relocked automatically, and the measurement is resumed.

\section{\label{sec:performance}Performance}

\begin{figure}[tbp]
\centering
\includegraphics{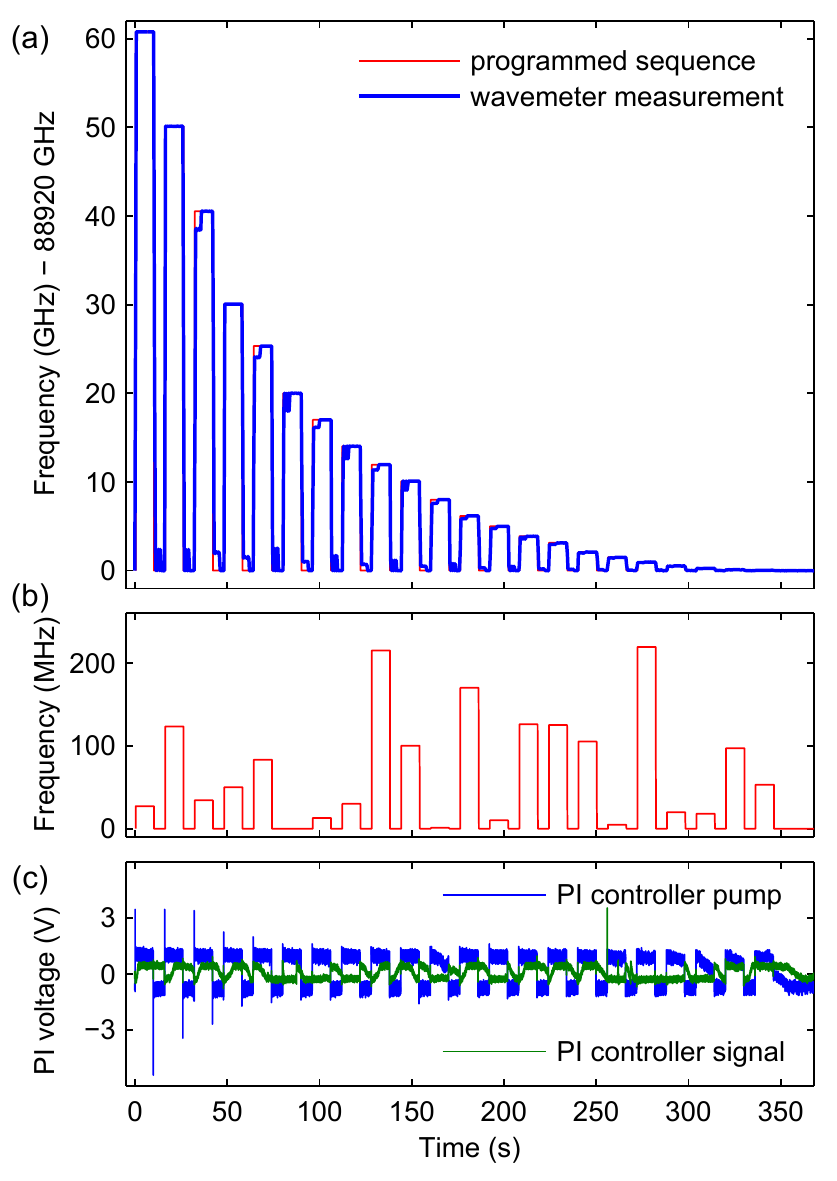}
\caption{
Sequence of frequency ramps with different amplitude. (a)~The programmed idler frequency compared to the wavemeter measurement. (b)~Programmed frequency modulo \unit[250]{MHz}. (c)~Measured PI controller voltage. The timescale of the sequence is chosen such that the wavemeter can measure the idler frequency sufficiently well.
}
\label{fig:rampseq}
\end{figure}

\begin{figure}[tbp]
\centering
\includegraphics{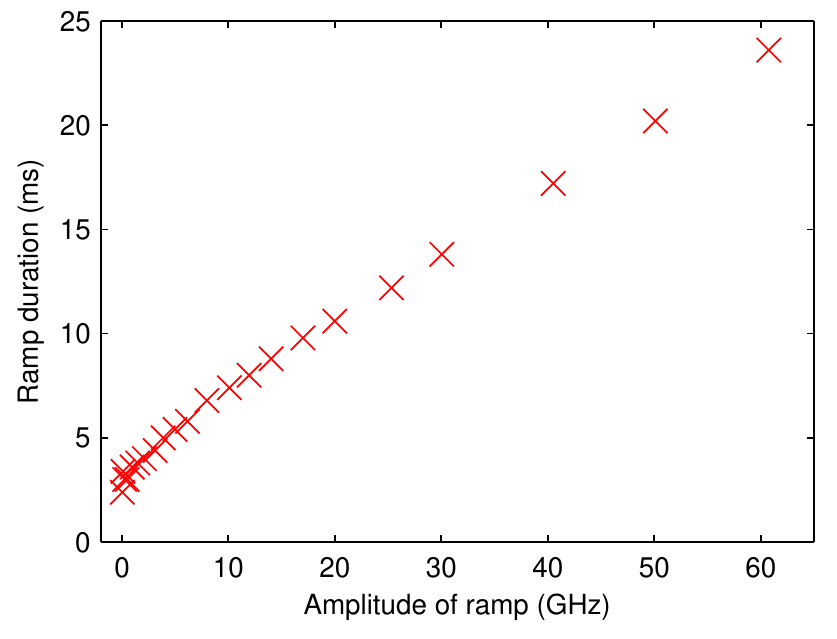}
\caption{
Measured ramp duration vs. amplitude of a frequency ramp.
}
\label{fig:ramptime}
\end{figure}

In the final section of the paper we want to discuss the performance of the system. In particular, we evaluate the accessible frequency range and the duration of frequency ramps with different amplitude.

Our approach of frequency control allows tuning over a broad range, limited by the mode-hop free tuning range of the laser. For our pump laser a tuning range of 100\,GHz is specified. However, we are limited to about 70\,GHz by our high-voltage amplifier at the moment. To demonstrate the tuning capability we perform a sequence of frequency ramps spanning tens of MHz up to more than 60\,GHz. The data is displayed in Fig.~\ref{fig:rampseq} and consists of ramps to 22 target idler frequencies and back to the initial value of 88920\,GHz. Here, the frequency is verified by reading out the wavemeter after every ramp. The programmed sequence and the measured frequency are shown in Fig.~\ref{fig:rampseq}(a). The wavemeter sometimes needs up to 2\,s to adjust to the new frequency and display a correct result. Apart from that we observe a perfect overlap. To demonstrate not only wide ramps but also precise locking anywhere inside the tuning range we chose target idler frequencies with random offsets to the OFC. As these offsets are not visible in part (a), we show them in Fig.~\ref{fig:rampseq}(b) where we plot the target frequencies modulo $f_{rep}=250\,\mathrm{MHz}$. To verify the instant relock to the OFC after each ramp we recorded the output voltages of the PI controllers with an oscilloscope (Fig.~\ref{fig:rampseq}(c)).

Figure~\ref{fig:ramptime} shows the ramp duration vs. the amplitude of a ramp. The time was measured with an oscilloscope during the sequence discussed in the previous paragraph. For ramps spanning more than 2\,GHz we observe a ramp speed of 2.9\,GHz/ms plus about 3\,ms offset for the slowdown of the ramp and the relock to the OFC. Shorter ramps are performed in about 3\,ms. With these values we did not reach the maximum tuning speed of our pump laser yet. We are rather limited by the repetition rate of the frequency counter, because we need a couple of frequency measurements on each slope of the beat frequency tooth structure (see Fig.~\ref{fig:rampanalysis}(a)) for correct frequency tracking. Consequently, faster tuning could be easily achieved with faster electronics. We note that the time needed for short frequency ramps could be lowered substantially with some further optimization of ramp parameters.

\section{\label{sec:conclusion}Conclusion}

We have demonstrated a method for frequency control of lasers referenced to an optical frequency comb. Our approach allows for fast and wide tuning of the laser over many GHz and within milliseconds and grants frequency stabilization to sub-MHz precision. We implemented the technique for an OPO, but it is suitable for any widely tunable continuous wave laser which can be referenced to an OFC. The frequencies of the OFC are not manipulated and, thus, the OFC can be shared among different laboratories and for control of many lasers.

We developed the system and optimized the ramping algorithm in order to address several rovibrational transitions in trapped molecules in the same experimental sequence and with a single OPO. Frequency ramping over many GHz on a millisecond timescale allows us to address molecules in many rotational states shortly after each other. We use this to control the internal molecular state by performing optical pumping on several transitions quasi-simultaneously. Flexible frequency tuning additionally permits us to use one OPO as a multi-purpose laser performing cooling, state preparation, and depletion spectroscopy in the same run of an experiment\cite{Glockner2015a}. Using the system described here, we demonstrated rotational cooling of methyl fluoride (CH$_3$F)~\cite{Glockner2015a} (idler wave of OPO at wavelength of 3.4\,\textmu m) and optoelectrical cooling\cite{Zeppenfeld2009} of formaldehyde (H$_2$CO) to the ultracold regime\cite{Prehn2016} (idler wave at 3.6\,\textmu m).

To facilitate experiments, the software running on the microcontroller can perform sequences of ramps and stabilize the OPO to many different frequencies autonomously. Additionally, the OPO can be controlled (within its electrical tuning range) remotely from a computer via the microcontroller. Our control software also relocks the OPO automatically whenever an error is detected and resumes any running experimental sequence. This allows for around-the-clock, hands-free measurements in the present setup. With a properly aligned system, we routinely perform more than $10^6$ frequency ramps without a single error.

\end{document}